\documentclass[twocolumn,showpacs,preprintnumbers,amsmath,amssymb,prl,floatfix]{revtex4}
\usepackage{psfig}
\usepackage{epsfig}
\expandafter\ifx\csname package@font\endcsname\relax\else
 \expandafter\expandafter
 \expandafter\usepackage
 \expandafter\expandafter
 \expandafter{\csname package@font\endcsname}%
\fi
\begin{document}

%\preprint{APS/123-QED}

\title{Magnetocapacitance:  A probe of spin-dependent potentials\\}% Force line breaks with \\

\author{K. T. McCarthy}
\author{A. F. Hebard}%
 \email{afh@phys.ufl.edu}
\affiliation{%
Department of Physics, University of Florida, Gainesville, FL 32611-8440\\
}%
\author{S. B. Arnason}%
\affiliation{%
Department of Physics, University of Massachussetts, Boston, MA 02125\\
}%
%\homepage{http://www.Second.institution.edu/~Charlie.Author}

\date{\today}% It is always \today, today,
             %  but any date may be explicitly specified

\begin{abstract}
The magnetic field dependence of the capacitance of Pd-AlO$_x$-Al thin-film
structures has been measured.  The observed quadratic dependence of capacitance on
magnetic field is consistent with a theoretical model that includes the effect of a spin-dependent
electrochemical potential on electron screening in the paramagnetic Pd. This
spin-dependent electrochemical potential is related to the Zeeman splitting of the narrow
d-bands in Pd.  The quantitative details depend on the electronic band structure at the
surface of Pd.
\end{abstract}

% PACS, the Physics and Astronomy
                             % Classification Scheme.
%\keywords{diluted magnetic semiconductors}%Use showkeys class option if keyword
                              %display desired
\maketitle

The design and implementation of spintronic devices demands accurate
experimental characterization of magnetic metals and semiconductors.  Specifically, a
detailed understanding of metal-semiconductor and metal-dielectric interfaces is
necessary.  Three examples illustrate this point:  efficient spin injection from a
ferromagnetic metal into a semiconductor
depends critically on interface properties\cite{Ham99203,Smi01045323};
the spin polarization of the tunneling
current in magnetic tunnel junctions is affected by spin-dependent surface
screening in the ferromagnetic electrodes\cite{Zha99640}; and
the magnitude of the magnetoresistance in
GMR is determined in part by the quality of the interfaces between ferromagnetic and
nonmagnetic layers\cite{Ben97288}.  The study of these and other interface effects
facilitates a better
understanding of the relevant physics in these devices.

%%%%%%%%%%%%%%%%%%%%%%%%%%%%%%%%%%%%%%%%%%%%%%%%%%%%%%%%%%%%%
\begin{figure}[b]
%\centerline{\fbox{\includegraphics{fig1.eps}}}% Here is how to import EPS art
\begin{center}
      \includegraphics[width=0.9\linewidth]{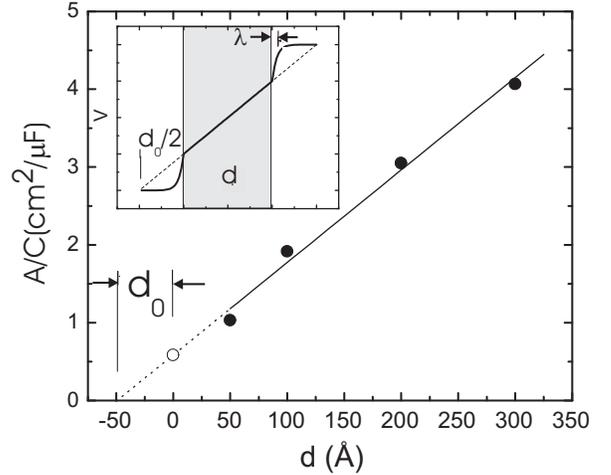}
\caption{\label{fig:fig1}
Inverse areal capacitance versus dielectric thickness for four Pd-AlO$_x(d)$-Al
structures at 10K. The negative of the $x$-intercept gives the additional effective thickness
$d_0$, which is 49\AA\ from a linear fit.  Similarly, the inverse slope determines that the
permittivity $\kappa$ of our AlO$_x$ is 9.5.  The inset illustrates our model of potential
versus position in which two metallic electrodes are separated by a dielectric (shaded
region) of thickness $d$.  The screening length is given by $\lambda$.
}
\end{center}
\end{figure}
%%%%%%%%%%%%%%%%%%%%%%%%%%%%%%%%%%%%%%%%

This project is motivated by Zhang's theoretical work\cite{Zha99640} involving electron screening in
ferromagnets.  According to Zhang, both Coulomb and exchange interactions influence the screening
response.  This fact is represented theoretically by a spin-dependent
potential decaying exponentially into the surface of a ferromagnet in the
presence of an electric field applied perpendicular to the surface.  The
spin-dependence originates from the exchange
splitting of spin-up and spin-down bands.
In this letter, we present the magnetic field
dependence of the screening length $\lambda$ of paramagnetic Pd which arises from a
magnetic field induced spin splitting of the density states.
This spin dependence of the screening potential in Pd is captured in magnetocapacitance measurements.
Our technique demonstrates a static probe of magnetic field-induced spin-dependent potentials
involving a voltage and no current.  This is in contrast to previous
work\cite{Ham99203,Smi01045323,Ben97288,Joh932142} involving
current-driven spin-accumulation.

When a capacitor is fabricated with a thin-film spacer layer,
a significant portion of the potential drop can occur across the
metal-insulator interfaces\cite{Mea61545,Sim6554}. This is shown schematically in the Fig. 1 inset.
The applied voltage $V$, the dielectric constant
$\kappa$, the dielectric thickness $d$, and the screening length $\lambda$ determine
the magnitude of these interfacial voltage drops.
The measured capacitance of a thin metal-insulating-metal
structure is thus indicative of bulk dielectric as well as interfacial
properties that include dependence on $\lambda$\cite{Mea61545,McC99302,Kru90199,Heb871349}.
The variation of screening length with magnetic field results in
a magnetic field dependent capacitance.
We chose Pd as our model system because of its large Pauli paramagnetic
susceptibility.  We measured the magnetic field dependent capacitance (magnetocapacitance)
of Pd-AlO$_x$-Al thin film structures and determined that the screening length of Pd increases
quadratically with applied magnetic field.

Our structures are grown and characterized in the following way.
First we use dc-magnetron sputtering to deposit 1000\AA\ of
Pd onto the entire surface of a square silicon substrate (1 cm$^2$).
A calibrated thickness of AlO$_x$ is then
grown over the entire Pd surface via reactive ion beam sputter deposition of Al
in an oxygen ambient at a carefully controlled pressure.
This technique has been shown\cite{Heb871349}
to produce dense, high quality, amorphous AlO$_x$.
The final deposition step utilizes a
shadow mask with 1mm diameter circular holes placed in close proximity to the sample
through which Al is thermally grown.

The sample is inserted into a Quantum Design
QD6000 cryostat in a custom sample probe with coaxial electrical leads attached to the
sample.  The bottom of the sample probe is an electrically isolated shielded enclosure
that houses the sample.
Three terminal capacitance measurements are
made using an AH-2700 capacitance bridge.  All measurements are performed in an
electrically screened room.  Capacitance was measured at a frequency of 1kHz and at
temperatures and magnetic fields ranging from 300K to 10K and -7T to 7T respectively.

Figure 1 shows the inverse areal capacitance versus dielectric
thickness at 10K for four Pd-AlO$_x$-Al structures with dielectric thickness ranging from
50\AA\ to 300\AA.
The linear dependence of inverse capacitance on thickness with nonzero intercept implies
that the geometrical
capacitance $C_g$ is in series with a thickness independent interface capacitance $C_i$
determined largely by the screening lengths in the two electrodes.
Other contributions to $C_i$, such as interface states and surface roughness, are present.
These additional contributions give rise to enhanced dispersion at low
frequency\cite{McC99302}. 
By plotting inverse capacitance versus dielectric thickness, we are able to infer the magnitude of
the interface capacitance, $C_i$, and an effective length scale, $d_0$, which defines the
crossover plate separation below (above) which $C_i$ ($C_g$) dominates.
$C_i$ is equal to the reciprocal of
the $y$ intercept (open circle) and $d_0$ is the x-intercept.
As $d$ approaches zero (and $C_g$ diverges) the measured capacitance
$C_m$ approaches $C_i$.  Our simple model for the capacitance of our structures is shown
schematically in the inset of Fig.~1 where the electrostatic potential, which decays
exponentially into each electrode, is plotted versus position.  For such a structure, $C_m$ is
given by
%%%%%%
\begin{equation}
\frac{A}{C_m}=\frac{A}{C_i}+\frac{A}{C_g}=\frac{d}{\kappa \epsilon_0}+\frac{d_0}{\kappa \epsilon_0}
\label{eq:one}
\end{equation}
%%%%%%
Where $d_0 = \kappa (\lambda_1 + \lambda_2)$.
The area is given by $A$, $\lambda_1$ and $\lambda_2$ are the screening lengths in the
metallic electrodes, and $\epsilon_0$
is the permittivity of free space. This model accurately accounts for our observed
thickness dependence.
It also shows us how capacitance measurements are
sensitive to the screening lengths in the electrodes of capacitors having sufficiently
thin spacer layers ($d \approx d_0$).
%%%%%%%%%%%%%%%%%%%%%%%%%%%%%%%%%%%%%%%%%%%%%%%%%%%%%%%%%%%%%
\begin{figure}[t]
%\centerline{\fbox{\includegraphics{fig2.eps}}}% Here is how to import EPS art
\begin{center}
      \includegraphics[width=0.9\linewidth]{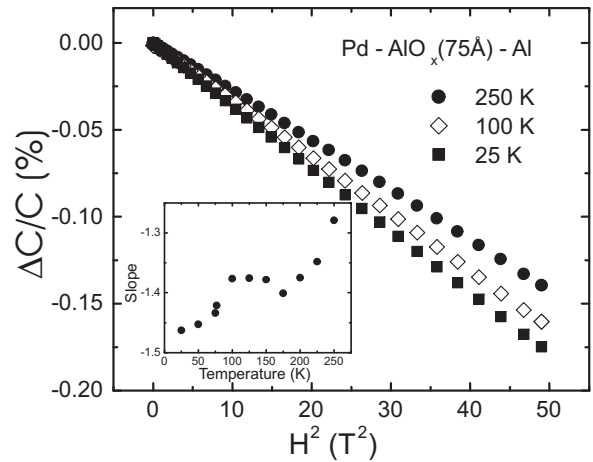}
\caption{\label{fig:fig2}
Plot of $100*(C(H) - C(0)/C(0))$ versus $H^2$ for Pd-AlO$_x$ (75\AA)-Al at the three
temperatures shown in the legend.  The slopes of these curves are plotted versus
temperature in the inset.
}
\end{center}
\end{figure}
%%%%%%%%%%%%%%%%%%%%%%%%%%%%%%%%%%%%%%%%

We observe that the capacitance of our structures decreases quadratically with
magnetic field (indicating a quadratic increase in $\lambda$) as shown in Fig.~2.
In this plot, the
normalized change in capacitance versus $H^2$ is linear with a negative slope.  The weak
temperature dependence of the slope is shown in the inset.  Strain effects can be ignored
since the linear magnetostriction in Pd films is zero to within one part
in 10$^6$\cite{Sch841073}.

By observing negative magnetocapacitance, we are inferring the $H$ dependence of
the screening length of Pd, since $\lambda$ is the only quantity depending directly on field.
The $H$ dependence comes from the field induced spin splitting of the conduction band of Pd.
Using a Stoner model of electron-electron interactions, the spin-dependent energy of an
electron is given by
%%%%%%
\begin{equation}
\epsilon^{\uparrow,\downarrow}=\epsilon(\textbf{k})+U_0 n^{\downarrow, \uparrow} \mp \mu_BH-e\phi(x)
\label{eq:two}
\end{equation}
%%%%%%
where $U_0$ is the energy that determines the strength of the interactions, $\epsilon(\textbf{k})$ is the band
energy, $\mu$ is the chemical potential, $\phi(x)$ is the electrostatic potential, $e$ is the
magnitude of the charge of the electron, $\mu_B$ is the Bohr magneton, and the spin-up (-down) carriers
are determined by the minus (plus) sign preceding the Zeeman energy ($\mu_B H$).  Further, the
spin-dependent carrier concentration at the surface of a metal with an electric field applied
normally (at $T=0$) is given by
%%%%%%
\begin{equation}
n^{\uparrow,\downarrow}(x)=\frac{1}{2} \int_{0}^{\mu-U_0 n^{\downarrow,\uparrow} \pm \mu_BH+e\phi(x)}
N(\epsilon) d\epsilon
\label{eq:three}
\end{equation}
%%%%%%
where $N(\epsilon)$ is the density of electronic states.
If we expand the density of states around the Fermi level, keep only the lowest order terms in $H$
and $\phi(x)$, and require charge conservation in the absence of an electric field
(to find the $H$ dependence of the chemical potential, $\mu$) we find that the total carrier
concentration a distance $x$ into the surface of the metal is given by
%%%%%%
\begin{eqnarray}
n^{\uparrow}(x)+n^{\downarrow}(x)=n_0+N(\epsilon_F)
\left(\frac{1}{1+J}\right)
\Biggl[ 1+\frac{1}{2}\left(\frac{1}{1+J}\right)\nonumber\\
\left( \eta+\frac{2J}{1-J}\left[ \frac{N'}{N} \right]^2_{\epsilon_F}\right)
\left(\frac{\mu_BH}{1-J}\right)^2 \Biggr] e\phi(x)\hspace{10mm}
\label{eq:four}
\end{eqnarray}
%%%%%%
with
%%%%%%
\begin{equation}
\eta = \left[\frac{N^{''}}{N}-{\left(\frac{N^{'}}{N}\right)}^2\right]_{\epsilon_F}
\label{eq:five}
\end{equation}
%%%%%%
and $J=\frac{1}{2}U_0N(\epsilon_F)$.
Poisson's equation must be satisfied self consistently, yielding an exponentially decaying
solution for the electrostatic potential over a length scale (screening length) given by
%%%%%%
\begin{eqnarray}
\lefteqn{\frac{1}{\lambda^2}=\frac{1}{\lambda_{TF}^2}\left(\frac{1}{1+J}\right)\Biggl[
1+\frac{1}{2}\left(\frac{1}{1+J}\right)} \nonumber\\
&& \left( \eta+\frac{2J}{1-J}
\left[\frac{N'}{N} \right]^2_{\epsilon_F}\right) \left(\frac{\mu_BH}{1-J}\right)^2 \Biggr]\hspace{5mm}
\label{eq:six}
\end{eqnarray}
%%%%%%
where $\lambda_{TF}^{-2}=e^2N(\epsilon_F)/\epsilon_0.$
All state densities and derivatives are evaluated at the Fermi energy $\epsilon_F$.
We note that the temperature and magnetic field dependence of the magnetic susceptibility of Pd are
treated similarly in the literature\cite{And70883,Bea805400,Mis711632}.
In fact, $\eta$ is the same band parameter that determines
the strength of the temperature dependence of the magnetic susceptibility of Pd at low
temperatures, namely:
%%%%%%
\begin{equation}
\chi_{T}=\chi_0\left[1+\frac{\pi^2}{6} \eta D(k_BT)^2\right]\:,
\label{eq:seven}
\end{equation}
%%%%%%
where $\chi_0 = \mu_{B}^{2} D N ( \epsilon_F )$ and $D=1/(1-J)$ is the Stoner enhancement
factor\cite{And70883}.
The Stoner-enhanced Pauli paramagnetic susceptibility is given by $\chi_0$ and we have
neglected terms of order $T^4$ and higher.

The capacitance of a structure containing one electrode with a magnetic field
dependent screening length as given by Eq.~6 (to quadratic order in $H$) is
%%%%%%
\begin{equation}
\frac{C(H)-C(0)}{C(0)}=\frac{C(0)}{A}\left[\frac{\lambda_{TF}}{4\epsilon_0}
\left( \frac{1}{1+J} \right)^{\frac{3}{2}} \gamma \left( \frac{\mu_BH}{1-J} \right)^2 \right]
\label{eq:eight}
\end{equation}
%%%%%%
where
%%%%%%
\begin{equation}
\gamma=\eta+\left(\frac{2J}{1-J}\right)\left[\frac{N'} {N}\right]^2_{\epsilon_F}\:.
\label{eq:nine}
\end{equation}
%%%%%%
Consolidating all of the $H$ dependence into the Pd electrode is justified by our observation
that magnetocapacitive effects in Al-AlO$_x$-Al structures are at least an order of
magnitude smaller than the same in Pd-AlO$_x$-Al structures.  Assuming that the screening lengths
of Al and Pd are similar, and using the values of $d_0$ and $\kappa$ determined from Fig.~1
(49\AA\ and 9.5 respectively), we estimate the $H=0$ screening length of Pd,
$\lambda_{TF}/(1+J)^{1/2}$, to be $\approx$~2.5\AA.  This length is artificially
large since it attributes all of the interface capacitance, $C_i$, to the screening response
of the electrodes, thereby neglecting surface roughness and interface states\cite{McC99302}.
Using the ($H = 0$) screening length, the slope of the $\Delta C/C$ vs $H^2$ curve and Eq.~8,
we estimate the strength of the field dependence, $\gamma$, to range from -100 to -1000 eV$^{-2}$.
This value varies from sample to sample, indicating the sensitivity of the magnetocapacitance
effect to interface quality.
Andersen\cite{And70883} performs a bulk calculation of $\eta$ for Pd using a rigid band
approximation.  He reports a value of 12 eV$^{-2}$, but it should be noted that the screening
length of a metal depends on its surface band structure, which may differ
substantially from the bulk band structure.
Eq.~9 implies that there is also an additional term that contributes to the strength of the magnetic
field dependence of the
screening length, which is absent in the temperature dependence of the magnetic
susceptibility ($\gamma>\eta$).  The presence of exchange interactions simply renormalizes the Zeeman
energy in the case of the susceptibility, as shown in Eq.~7 ($D\approx 10$)\cite{And70883}.

%%%%%%%%%%%%%%%%%%%%%%%%%%%%%%%%%%%%%%%%%%%%%%%%%%%%%%%%%%%%%
\begin{figure}[t]
%\centerline{\fbox{\includegraphics{fig3.eps}}}% Here is how to import EPS art
\begin{center}
      \includegraphics[width=0.9\linewidth]{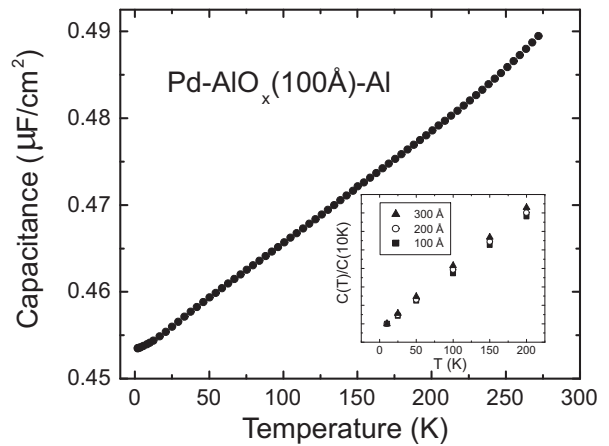}
\caption{\label{fig:fig3}
Plot of areal capacitance versus temperature for Pd-AlO$_x$(100\AA)-Al.  The inset
shows normalized capacitance versus temperature for structures with AlO$_x$ thicknesses
indicated in inset legend.
}
\end{center}
\end{figure}
%%%%%%%%%%%%%%%%%%%%%%%%%%%%%%%%%%%%%%%%

The chemical potential, $\mu$, in Pd lies so close to an inflection point in the density of states
versus energy that any perturbation resulting in a small change in $\mu$ can affect the magnitude
and sign of $\eta$.
We believe the chemical potential in the Pd electrode near the Pd-AlO$_x$ interface lies very near
a peak in the density of states, which accounts for our observation that $\gamma<0$
(therefore $\eta<0$) as determined by magnetocapacitance measurements.
Indeed it is predicted that doping Pd with additional d-electrons (e.g. alloying with Ag),
thereby modifying the chemical potential, causes the sign of $\eta$ to reverse\cite{And70883}.
The sign change in $\eta$ arises from a competition between the first and second derivatives of
the state densities at the Fermi level as indicated in Eq.~5.

The capacitance of our structures is temperature dependent, and decreases linearly
(at high temperatures) with decreasing temperature, as shown in Fig.~3.  The inset shows
a capacitance (normalized to 10K) versus temperature for three AlO$_x$ thicknesses and
indicates a trend toward stronger temperature dependence for thicker oxides. The first
correction to the Thomas-Fermi screening length due to temperature is quadratic;
therefore the operative mechanism behind the temperature dependence is not electronic in
nature except, perhaps, at the lowest temperatures.  We attribute the temperature
dependence in the linear regime to temperature-dependent strain in the metallic electrodes in
conjunction with a temperature-dependent dielectric constant, which accounts for our
observation of stronger temperature dependence for thicker oxides:  as the oxide
thickness is increased, the capacitance is more strongly dependent upon bulk dielectric properties. 

Other effects may contribute, such as the freezing out of electron traps at the interfaces and
reduced frequency dispersion as the temperature is lowered.  The negative magnetocapacitance becomes
larger in magnitude as the capacitance decreases (Fig.~2 inset, $y$-axis contains only
negative values).  This is curious as the magnetocapacitance should scale as shown in
Eq.~8 with the $H=0$ capacitance, $C(0)$.  Though we do not fully understand this dependence, one
possible contribution to this phenomenon is as follows:  since the freezing out of traps in the
interface removes a subset of localized charged levels,
a larger fraction of the net interfacial charge resides in the electrodes themselves, and thus
a larger potential drop is associated with electric field penetration into the electrodes.  Only
the induced charge within the Pd electrode contributes to the magnetocapacitance, therefore
we expect the magnetic field dependence to be stronger when, at lower temperatures,
a larger portion of the potential
drop occurs due to electric field penetration into the Pd electrode.
In addition, the magnetic susceptibility of palladium shows a very strong dependence on temperature
with a pronounced peak near 80K\cite{Mis711632}. As indicated in Eq.~7
the origin of this peak depends on the same band structure factor, $\eta$, that appears in our
magnetocapacitance calculations.
Accordingly, it is not surprising that we
observe a similar extremum in the slope of the $\Delta C/C$ vs $H^2$ curve (Fig.~2 inset).

In conclusion, we have measured the magnetic field dependence of the screening
length of paramagnetic Pd via magnetocapacitance.  We have proposed a model that
captures the essence of the relevant physics of this effect.  We have also shown that
magnetocapacitance measurements reveal surface band structure, which is distinct from
bulk band structure.  Since magnetocapacitance is sensitive to spin-dependent
electrochemical potentials, it may be a technique capable of measuring non-equilibrium
spin polarization in spin-injection devices.  Novel, low carrier density materials, e.g.
dilute magnetic semiconductors, may show similar magnetocapacitive effects larger in
magnitude due to longer screening lengths.

The authors would like to acknowledge useful discussions with Dmitrii Maslov and Pradeep Kumar.  This
work was supported by NSF DMR-0101856.
\bibliography{final}
\end{document}